\documentclass[a4paper,12pt]{article}
\usepackage{amssymb,latexsym}

\makeatletter \@addtoreset{equation}{section} \makeatother

\newcommand{\rE}{\mathrm{E}}

\newcommand{\rSL}{\mathrm{SL}}

\begin{document}

\begin{titlepage}

\thispagestyle{empty}

\begin{flushright}
\hfill{CERN-PH-TH/2005-058}\\
\hfill{UCLA/05/TEP/12}
\end{flushright}

\vspace{35pt}

\begin{center}{ \LARGE{\bf
$\rE_{7(7)}$ symmetry and dual gauge algebra\\[5mm]
of M--theory on a twisted seven--torus}} \vspace{60pt}

{\bf  R. D'Auria$^\bigstar$, S. Ferrara$^\dag$ and M. Trigiante$^\bigstar $}

\vspace{15pt}

$^\bigstar${\it Dipartimento di Fisica, Politecnico di Torino \\
C.so Duca degli Abruzzi, 24, I-10129 Torino, and\\
Istituto Nazionale di Fisica Nucleare, \\
Sezione di Torino,
Italy}\\[1mm] {E-mail: riccardo.dauria@polito.it,  mario.trigiante@polito.it}

$^\dag$ {\it CERN, Physics Department, CH 1211 Geneva 23, Switzerland\\ and\\ INFN, Laboratori
Nucleari di Frascati, Italy\\and\\
Department of Physics \& Astronomy, University of California, Los Angeles, CA}\\[1mm] {E-mail: Sergio.Ferrarara@cern.ch}

\vspace{50pt}

{ABSTRACT}

\end{center}

\medskip

We consider M--theory compactified on a twisted 7--torus with
fluxes when all the seven antisymmetric tensor fields in four
dimensions have been dualized into scalars and thus the
$\rE_{7(7)}$ symmetry is recovered. We find that the
Scherk--Schwarz and flux gaugings define a ``dual'' gauge algebra,
subalgbra of $\rE_{7(7)}$, where some of the generators are
associated with vector fields which are dual to part of the
original vector fields (deriving from the 3--form). In particular
they are dual to those vector fields which have been ``eaten'' by
the antisymmetric tensors in the original theory by the
(anti--)Higgs mechanism. The dual gauge algebra coincides with the
original gauge structure when the quotient with respect to these
dual (broken) gauge generators is taken. The particular example of
the S-S twist corresponding to a ``flat group'' is considered.

\end{titlepage}

\newpage

\section{M--theory on twisted tori and its gauge structure}
Compactification of superstring or M--theory on twisted tori \cite{ss,km,df,alt,ddf,h} in the
presence of form--fluxes \cite{svw,bb} offers interesting models where a scalar potential for
the moduli fields is obtained and gauge and supersymmetry breaking remove most (if not all)
the flat directions of the original geometry.\par From the point of view of the low--energy
effective theory in lower dimensions (four dimensions in our case) the change in the geometry
of the internal space results in a ``massive'' deformation of a certain supergravity theory,
where the scalar potential and the Yukawa couplings are induced by the underlying gauge
algebra structure of the gauged supergravity. The low--energy dynamics of M--theory
compactified on a 7--torus is described by an effective $D=4$, $N=8$ (maximal) supergravity.
We shall refer in the following to the maximal theory in four dimensions with 70 scalar fields
(and, in the absence of twists or fluxes, with a manifest $\rE_{7(7)}$ global invariance of
the combined equations of motion and Bianchi identities \cite{cj}) as \emph{standard}
$N=8,\,D=4$ supergravity \footnote{Note that this definition of $N=8, D=4$ standard
supergravity comprises ungauged theories which, in spite of the common $\rE_{7(7)}$ global
symmetry at the level of field equations and Bianchi identities, have inequivalent Lagrangians
with different global symmetry groups. These models therefore offer different choices of gauge
symmetry which can be introduced \cite{adfl,dwst}.}. In the case of M--theory an interesting
phenomenon occurs due to the fact that the ordinary compactification on a 7--torus with fluxes
results in an unconventional maximal supergravity theory in $D=4$ where 7 of the 70 scalar
fields have been replaced by antisymmetric tensors. The $\rSL(7,\mathbb{R})$ assignment of the
vectors and tensors from the M--theory compactification is: ${\bf 21}+\overline{{\bf 7}}$ for
the 28 vector fields and ${\bf 7}$ for the tensor fields. Denoting by $A_{IJ\mu},\,A^I_\mu$
(in our notations the capital Latin indices label the internal directions of the 7-torus:
$I,J\dots =1,\dots, 7$) the 1--form and by $B_{I\mu\nu}$ the 2--form fields, their combined
algebraic structure results in a ``free differential algebra'' (FDA) \cite{sullivan} which was
studied in \cite{ddf}. In particular the Scherk--Schwarz (S-S) structure constants
$\tau_{IJ}{}^K$ play the role, in this algebra, of a ``magnetic'' mass for the B--field. To
show this let us consider the general 2--form structure which appear in the low--energy
theory:
\begin{eqnarray}
F^\Lambda&=& dA^\Lambda+\frac{1}{2}\,f_{\Sigma\Gamma}{}^\Lambda\,A^\Sigma\,
A^\Gamma+m^{\Lambda I}\,B_I\,,\label{FL}
\end{eqnarray}
in which we have generically denoted by an upper index $\Lambda$
the lower antisymmetric couple $[IJ]$:
\begin{eqnarray}
F^\Lambda&=& F_{IJ}\,,
\end{eqnarray}
and we have identified the $21\times 7$ matrix $m^{\Lambda I}$
with  $\tau_{KL}{}^I$. If we assume this matrix to have rank $r\le
7 $ then $r$ of the (redefined) 2--forms $B_I$, denoted by
$\hat{B}_\alpha$, appear in the following combination:
\begin{eqnarray}
F^\alpha & =& dA^\alpha+m^{\alpha\beta}\,
\hat{B}_\beta\,.\label{F}
\end{eqnarray}
In the vacuum $F^\alpha=0$ and therefore $d\hat{B}_\alpha=0$. Equation (\ref{F}) indicates
that the gauge fields $A^\alpha$ make the $\hat{B}_\alpha$ massive and that the quotient
algebra of the $A^\Lambda$ (with the $A^\alpha$ modded out) is an ordinary Lie algebra. The
connection $\Omega$ and structure (non vanishing commutators) of this algebra have the
following form \cite{df}:
\begin{eqnarray}
\Omega&=&A^I_\mu\,Z_I+A_{IJ\mu}\, W^{IJ}\,,\nonumber\\
\left[Z_I,\,Z_J\right]&=&\tau_{IJ}{}^K\,Z_K+g_{IJKL}\,
W^{KL}\,\,\,;\,\,\,\,\,\left[Z_I,\,W^{JK}\right]=
2\,\tau_{IL}{}^{[J}\, W^{K]L}\,,\label{modg}
\end{eqnarray}
(here $W^{IJ}$ are the remaining $21-r$ generators) with
\begin{eqnarray}
\tau_{[IJ}{}^P\,\tau_{K]P}{}^M&=&\tau_{[IJ}{}^P\,g_{KLR]P}=0\,.\label{constr}
\end{eqnarray}
The constraints (\ref{constr}) represent the Jacobi identities of the structure constants
$f_{\Sigma\Gamma}{}^\Lambda$ in eq.(\ref{FL}) as they come from the free differential algebra
\cite{ddf} closure condition or, equivalently, from the 4--form Bianchi identity in M--theory
\cite{df}. However condition (\ref{constr}) does not guarantee closure of the algebra
(\ref{modg}). The latter indeed turns out to close under the general conditions (\ref{constr})
only if regarded either as part of the FDA ({\ref{FL}) or, as we shall see, as part of a
larger gauge algebra in the standard formulation of the theory.
\par In the particular case in which the only non vanishing
entries of the $\tau_{IJ}{}^K$ and $g_{IJKL}$ tensors are $\tau_{1 i}{}^j= T_i{}^j$ (to be
considered as an invertible $6\times 6$ matrix) and $g_{1ijk}$ ($i,j=2,\dots, 7$), the
quotient algebra is a 22--dimensional Lie algebra spanned by $Z_1,\,Z_i,\,W_{ij}$ with
structure constants $T_i{}^j,\,g_{1ijk}$. In the framework of standard maximal supergravity,
the gauged theory corresponding to this choice of twist  $\tau$ and flux $g$ was shown in
\cite{alt} to coincide with the model originating from a $D=5\rightarrow D=4$ generalized S-S
dimensional reduction \cite{adfl,css,svn}, for a suitable choice of the S-S twist generator in
$\rE_{6(6)}$. If moreover $T_i{}^j=-T^j{}_i$ the gauge algebra defines a \emph{flat group} in
the language of \cite{ss} and the corresponding theory admits a Minkowski vacuum.
\par
Note that the gauge algebra (\ref{modg}) does not arise from a gauging of standard $N=8,\,D=4$
supergravity since, as we shall comment on later, it is not contained in the global symmetry
algebra of the theory. The unusual situation is due to the fact that in the presence of
antisymmetric tensor fields the original $\rE_{7(7)}$ global symmetry of $D=4,\,N=8$
supergravity is lost. In particular the $\rE_{7(7)}$ isometries corresponding to the 7 scalars
$\tilde{B}^I$ which have been dualized into antisymmetric tensor fields have been replaced by
tensor gauge symmetries with the implication that the generators $W^{IJ}$ corresponding to the
gauge fields $A_{IJ\mu}$ have become abelian, as in the algebra (\ref{modg}).
\par The present note is organized as follows: In section 2  we
consider the standard formulation of the four dimensional theory where the antisymmetric
tensor fields $B_{I\mu\nu}$ have been replaced by the scalars $\tilde{B}^I$ and thus the
$\rE_{7(7)}$ global symmetry of the equations of motion and Bianchi identities is manifest. In
this framework, we shall apply the general group theoretical analysis of gauged maximal
supergravities developed in \cite{dwst}, to the construction of the gauge algebra arising from
the presence of twist--tensor $\tau_{IJ}{}^K$ and flux $g_{IJKL}$. This approach is based on
the description of the most general gauging of the $D=4,N=8$ theory in terms of an
$\rE_{7(7)}$--covariant tensor $\theta$ called \emph{embedding matrix}, which defines the
gauge generators as linear combinations of the global symmetry generators. The advantage of
such description is that the $\rE_{7(7)}$ invariance of the combined field equations and
Bianchi identities of the gauged theory is restored provided $\theta$ is transformed together
with the fields of the model. Supersymmetry requires $\theta$ to transform in the ${\bf 912}$
and closure of the gauge algebra inside $\rE_{7(7)}$ implies further quadratic constraints in
$\theta$. In this formalism the background quantities $\tau_{IJ}{}^K$ and $g_{IJKL}$ are
identified with components of the embedding tensor. To be specific they correspond to the
${\bf 140}$ (considering $\tau$ to be traceless) and $\overline{{\bf 35}}$ in the branching of
the ${\bf 912}$ with respect to $\rSL(7,\mathbb{R})$. Group theory then determines the gauge
algebra structure and we find that the second order constraints on $\theta$ correspond
precisely with the quadratic constraints (\ref{constr}) on $\tau$ and $g$. As we shall see,
the gauge generators consist not just of $Z_I$ and $W^{IJ}$, but also of some ``dual''
generators $W_{IJ}$ to be gauged by the magnetic vector fields $\tilde{A}^{IJ}_\mu$ dual to
$A_{IJ\mu}$. This seems in contradiction with the notion that a gauging which involves
magnetic vector fields is inconsistent. Actually what the presence of ``magnetic'' generators
$W_{IJ}$ is telling us is that the electric fields to be described locally in the low--energy
gauged theory are not $A^I_\mu,\,A_{IJ\mu}$, but are defined after a  symplectic rotation
between $A_{IJ\mu}$ and $\tilde{A}^{IJ}_\mu$ (whose existence is guaranteed by the second
order constraints \cite{dwst2}). This mechanism is described in detail in the case $\tau\equiv
\{\tau_{1i}{}^j\}$ and $g\equiv \{g_{1ijk}\}$, in which we show that this symplectic rotation
yields
 electric vector fields transforming in the representation $\overline{{\bf 27}}+{{\bf 1}}$
 of $\rE_{6(6)}$, thus confirming that the corresponding model can be
 alternatively obtained through a suitable $D=5\rightarrow D=4$
 generalized S-S reduction.\par
 In section 3 we review the FDA analysis of \cite{ddf}. Finally we
 shall end with some concluding remarks.

\section{Gaugings from M--theory with $\rE_{7(7)}$ symmetry}

We devote the present section to the derivation, using the
embedding tensor approach, of the gauge algebra structure in the
$D=4,N=8$ theory which originates from the presence a non trivial
twist tensor $\tau_{MN}{}^P$ and 4-form flux $g_{MNPQ}$. We shall
also consider the effect in terms of local symmetries of a flux
over the volume of $T^7$ of the 7-form $\tilde{g}$ dual to $g$.
The main result is the algebra (\ref{comms2}). This analysis
parallels the one of \cite{ddf} in which the antisymmetric tensors
are left undualized and the gauge structure originating from the
compactification is encoded in a free differential algebra, to be
reviewed in section 3.\par In standard four dimensional maximal
supergravity the electric and magnetic charges transform in the
${\bf 56}$ of $\rE_{7(7)}$ and, as anticipated in the
introduction, the most general gauging can be described in terms
of an embedding tensor \cite{dwst} $\theta_u{}^\sigma$
($u=1,\dots,56$ and $\sigma=1,\dots, 133$), which expresses the
generators $X_u$ of the gauge algebra ${\frak g}$ in terms of
$\rE_{7(7)}$ generators $t_\sigma$:
\begin{eqnarray}
X_u&=&\theta_u{}^\sigma\,t_\sigma\,.
\end{eqnarray}
In this notation, since the index $u$ runs over both electric and magnetic charges,
consistency of the gauging requires the rank of $\theta$ to be not greater than $28$.
Supersymmetry and closure of the gauge algebra inside $\rE_{7(7)}$ require a linear and a
quadratic condition in $\theta$ respectively \cite{dwst,dwst2}:
\begin{eqnarray}
\theta &\in& {\bf 912}\,\subset {\bf 56}\times {\bf 133}\,,\label{912}\\
&&\theta_u{}^\sigma\,\theta_v{}^\gamma\,\mathbb{C}^{uv}=0\,,\label{2c}
\end{eqnarray}
where $\mathbb{C}^{uv}$ is the ${\rm
Sp}(56,\mathbb{R})$--invariant matrix. The last condition ensures
that \emph{there always exists a symplectic rotation acting on the
index $u$ as a consequence of which all the vectors associated
with the generators $X_u$ are electric (or all magnetic).}\par
 Let us start considering the
branchings of the relevant $\rE_{7(7)}$ representations with respect to the subgroup ${\rm
SL}(7,\mathbb{R})\times {\rm O}(1,1)_2$ (the subscript $2$ will be used later to distinguish
the corresponding grading from the charge with respect to a different ${\rm O}(1,1)$ group):
\begin{eqnarray} {\bf 56}&\rightarrow & \overline{{\bf 7}}_{-3}+{\bf
21}_{-1}+\overline{{\bf 21}}_{+1}+{\bf 7}_{+3}\,,\nonumber\\
{\bf 133}&\rightarrow & {\bf 7}_{-4}+ \overline{{\bf
7}}_{+4}+\overline{{\bf 35}}_{-2}+{\bf 35}_{+2}+{\bf 48}_0+{\bf
1}_0\,,\nonumber\\
{\bf 912}&\rightarrow &{\bf 1}_{-7}+{\bf 1}_{+7}+{\bf
35}_{-5}+\overline{{\bf 35}}_{+5}+ (\overline{{\bf
140}}+\overline{{\bf 7}})_{-3}+ ({\bf 140}+{\bf 7})_{+3}+{\bf
21}_{-1}+\overline{{\bf 21}}_{+1}+\nonumber\\&&{\bf
28}_{-1}+\overline{{\bf 28}}_{+1}+{\bf 224}_{-1}+\overline{{\bf
224}}_{+1}\,.\label{branchs}
\end{eqnarray}
In the branching of the ${\bf 56}$ the $\overline{{\bf 7}}_{-3}$ and ${\bf 21}_{-1}$ define
$A^I_\mu,\,A_{IJ\mu}$ respectively while ${\bf 7}_{+3}$ and $\overline{{\bf 21}}_{+1}$ their
magnetic duals. In the branching of the adjoint representation of $\rE_{7(7)}$
 we denote by $t_M{}^N,\,t^{MNP},\,t_P$ the generators in the ${\bf 48}_0,\,{\bf 35}_{+2}$
 and $\overline{{\bf
7}}_{+4}$ respectively (with an abuse of notation we characterize
each generator by the representation of the corresponding
parameter, this allows a simpler interpretation of the table
below). The commutation relations between these generators is:
\begin{eqnarray}
\left[t_M{}^N,\,t^{PQR}\right]&=&-3\,\delta_M^{[P}\,t^{QR]N}\,\,;\,\,\,\left[t_M{}^N,\,t_{P}\right]=\delta_{P}^{N}\,t_{M}\,,\nonumber\\
\left[t_M{}^N,\,t_P{}^Q\right]&=&\delta^N_P\,t_M{}^Q-\delta^Q_M\,t_P{}^N\,\,;\,\,\,\left[t^{M_1
M_2M_3 },\,t^{M_4 M_5 M_6}\right]=\epsilon^{M_1\dots M_6
M}\,t_M\,.\label{come7}
\end{eqnarray}
 Each representation in the branching of ${\bf 912}$
defines
 a different set of entries of $\theta$ which can be switched on
independently of the others and leads to a specific gauging. It is
useful to arrange the above representations in a table as follows:
\begin{center}
{\small \begin{tabular}{|c||c|c|c|c|}\hline
    & ${\bf 7}_{+3}$ & $\overline{{\bf 21}}_{+1}$ & ${\bf
21}_{-1}$ & $ \overline{{\bf 7}}_{-3}$  \\\hline\hline
  $\overline{{\bf 7}}_{+4}$ & ${\bf 1}$ & $\overline{{\bf 35}}$ & ${\bf 140}+{\bf 7}$ &
   $\overline{{\bf 28}}+\overline{{\bf 21}}$ \\
  ${\bf 35}_{+2}$ & $\overline{{\bf 35}}$ &${\bf 140}$ & $\overline{{\bf 21}}+\overline{{\bf 224}}$ &
  ${{\bf 21}}+{{\bf 224}}$ \\
  ${\bf 48}_0$ & ${\bf 140}+{\bf 7}$ & $\overline{{\bf 21}}+\overline{{\bf 28}}+\overline{{\bf 224}}$ & ${{\bf 21}}+{{\bf 28}}+{{\bf 224}}$ & $\overline{{\bf 140}}+\overline{{\bf 7}}$\\
  ${\bf 1}_0$ & ${\bf 7}$ & $\overline{{\bf 21}}$ & ${\bf
21}$ & $ \overline{{\bf 7}}$ \\
  $\overline{{\bf 35}}_{-2}$ & $\overline{{\bf 21}}+\overline{{\bf 224}}$ & ${{\bf 21}}+{{\bf 224}}$ &
  $\overline{{\bf 140}}$ & ${\bf
35}$ \\
  ${\bf 7}_{-4}$ & ${{\bf 28}}+{{\bf 21}}$ & $\overline{{\bf 140}}+\overline{{\bf 7}}$ & ${\bf
35}$ & ${\bf 1}$ \\ \hline
\end{tabular}}
\end{center}
The first row and column contain the representations in the branchings of ${\bf 56}$ and the
${\bf 133}$ respectively, while the bulk contains representations in the branching of ${\bf
912}$. The table specifies the origin of the latter representations in the branching of the
product ${\bf 56}\times{\bf 133}$ and  it should be read as ``first row times first column
gives bulk''. The grading of each entry of the table has been suppressed for the sake of
simplicity, since it coincides with the sum of the gradings of the corresponding elements in
the first row and column.\footnote{In principle there could have been a representation ${\bf
7}_{+3}$ in the slot ${\bf 35}_{+2}\times
 \overline{{\bf 21}}_{+1}$ and a $\overline{{\bf 7}}_{-3}$ in the slot
 $\overline{{\bf 35}}_{-2}\times {\bf 21}_{-1}$. However the presence of these representations
 would be inconsistent with the corresponding table of branchings with respect to the ${\rm SL}(8,\mathbb{R})$
 maximal subgroup of ${\rm E}_{7(7)}$, which contains ${\rm GL}(7,\mathbb{R})$ (see table (6.2) of \cite{dwst}). }
\par To see what information can be gained from this table let us
choose to restrict ourselves to the components of $\theta$ in the
representations ${\bf 140}_{+3}$, $\overline{{\bf 35}}_{+5}$ and
${\bf 1}_{+7}$ contained in the ${\bf 912}$. The first two
correspond (modulo multiplicative factors) to the tensors
$\tau_{MN}{}^P$ and $g_{MNPQ}$ respectively, while the third is
related to the flux of the dual 7-form $\tilde{g}$ over $T^7$:
\begin{eqnarray}
{\bf 140}_{+3}&\leftrightarrow &\tau_{MN}{}^P\,,\nonumber\\
\overline{{\bf 35}}_{+5}&\leftrightarrow &g_{MNPQ}\,,\nonumber\\
{\bf 1}_{+7}&\leftrightarrow & \tilde{g}_{M_1\dots
M_7}=\tilde{g}\,\epsilon_{M_1\dots M_7}\,.
\end{eqnarray}
 We have a component of the embedding
tensor, depending only on $\tau$, which intertwines between the
electric charge in ${\bf 21}_{-1}$ and the $\rE_{7(7)}$ generator
$t_M$ in the $\overline{{\bf 7}}_{+4}$. This defines the following
first set of gauge generators:
\begin{eqnarray}
W_{MN}&=&\theta_{MN,}{}^P\,t_P=c_1\,\tau_{MN}{}^P\,t_P\,.\label{Wdd}
\end{eqnarray}
Note that there are at most 7 independent $W_{IJ}$ depending on the rank $r\le 7$ of the
$21\times 7$ matrix $\tau_{IJ}{}^K$. The we have two components
$\theta^{MN,}{}_{PQR},\,\theta^{MN,\,P}$ of the embedding tensor with the electric index $u$
in the same representation $\overline{{\bf 21}}_{+1}$.  They contract with the $\rE_{7(7)}$
generators $t^{PQR}$ in the  ${\bf 35}_{+2}$ through the tensor $\tau$ and with the generators
$t_P$ in the $\overline{{\bf 7}}_{+4}$ through $g$. These components define the following
generators:
\begin{eqnarray}
W^{MN}&=&\theta^{MN,}{}_{PQR}\,t^{PQR}+
\theta^{MN,\,P}\,t_P=\,b_1\,
\tau_{PQ}^{[M}\,t^{N]PQ}+b_2\,\epsilon^{MNM_1\dots M_4
P}\,g_{M_1\dots M_4 }\,t_P\,.\label{Wuu}
\end{eqnarray}
Finally there are three more components of $\theta$ which
intertwine between the representations ${\bf 7}_{+3}\in {\bf 56}$
and ${\bf 48}_{0}\in {\bf 133}$ through the tensor $\tau$, between
the representations ${\bf 7}_{+3}\in {\bf 56}$ and ${\bf
35}_{+2}\in {\bf 133}$ through the tensor $g$ and between the
${\bf 7}_{+3}\in {\bf 56}$ and the $\overline{{\bf 7}}_{+4}\in
{\bf 133}$ through the tensor $\tilde{g}$. They define the last
set of gauge generators:
\begin{eqnarray}
Z_M&=&
\theta_{M,\,M_1M_2M_3}\,t^{M_1M_2M_3}+\theta_{M,\,N}{}^P\,t_{P}{}^N+\theta_{M,}{}^N\,t_N=
a_1\,g_{M M_1M_2M_3}\,t^{M_1M_2M_3}+\nonumber\\&&a_2
\tau_{MN}{}^P\,t_{P}{}^N+a_3\,\tilde{g}\,t_M\,.\nonumber\\&&
\end{eqnarray}
The constraints (\ref{2c}) imply the following conditions:
\begin{eqnarray}
\theta_{MN,}{}^P\,\theta^{MN,}{}_{RST}&=&0\,\,\Rightarrow\,\,\,\tau_{[MN}{}^P\,\tau_{Q]P}{}^R=0\,\nonumber\\
\theta_{MN,}{}^P\,\theta^{MN,\,Q}-\theta_{MN,}{}^Q\,\theta^{MN,\,P}&=&0\,\,\Rightarrow\,\,\,
\tau_{[MN}{}^P\,g_{M_1M_2M_3]P}=0\,.\label{2ord}
\end{eqnarray}
Taking into account equations (\ref{Wdd}) and (\ref{Wuu}), we see that the generators $W^{MN}$
and $W_{MN}$ are not linearly independent since they satisfy the two constraints:
\begin{eqnarray}
\tau_{[PQ}}{}^N\,W_{R]N&=&0\,\nonumber\\
b_2\,\epsilon^{M_1M_2M_3M_4PQR}\,g_{M_1M_2M_3M_4}W_{QR}&=&c_1\,\tau_{ST}{}^PW^{ST}\,.\label{2cons}
\end{eqnarray}
 In particular this means that if $r$ is the rank of the $21\times 7$ matrix $\tau_{MN}{}^P$
 only $r$ $W_{PQ}$ generators and $(21-r)$  $W^{PQ}$ generators are linearly independent.\\
 The previous
analysis indicates that the gauge connection has the following form:
\begin{eqnarray}
\Omega_{\frak g\,\mu}
&=&A^M_\mu\,Z_M+A_{MN\mu}\,W^{MN}+\tilde{A}^{MN}_\mu\,W_{MN}\,.\label{conng}
\end{eqnarray}
where the restrictions (\ref{2cons}) are understood. Although this gauging involves the
vectors $A_{MN\mu}$ together with their duals $\tilde{A}^{MN}_\mu$, the conditions
(\ref{2cons})and (\ref{2ord}) guarantee that no more than 28 independent linear combinations
of them can take part to the minimal couplings, namely that there exists a symplectic frame in
which all the $A_{MN\mu}$ and $\tilde{A}^{MN}_\mu$ involved in this gauging are electric. An
other way of stating this is that \emph{ the gauging chooses its own symplectic frame}. This
symplectic frame is in general different from the ${\rm GL}(7,\mathbb{R})$ one in which the
magnetic charges (vector fields) transform in the $\overline{{\bf 7}}_{-3}+{\bf 21}_{-1}$. For
instance in the case to be discussed in the next subsection the natural frame is the
$\rE_{6(6)}\times{\rm O}(1,1)$ and the corresponding gauging coincides with the one
originating from  $D=5\rightarrow D=4$ S-S reduction \cite{css}, for a suitable choice of
parameters.\par The general structure of the algebra is:
\begin{eqnarray}
\left[Z_M,\,Z_N\right]&=&\alpha\,\tau_{MN}{}^P\,Z_P+\beta\,g_{MNPQ}\,W^{PQ}+\rho\, \tilde{g}\,W_{MN}\,,\nonumber\\
\left[Z_M,\,W^{PQ}\right]&=&\gamma\,\tau_{MR}{}^{[P}\,W^{Q]R}+\sigma\,g_{MM_1M_2M_3}\,\epsilon^{M_1M_2M_3PQRS}\,W_{RS}\,,\nonumber\\
\left[Z_M,\,W_{PQ}\right]&=&
\delta\,\tau_{PQ}{}^L\,W_{ML}\,\nonumber\\
\left[W^{IJ},\,W^{KL}\right]&=&-\frac{\lambda}{2}\,
\tau_{I_1I_2}{}^{[K}\,W_{I_3I_4}\epsilon^{L]IJI_1\dots
I_4}\,,\nonumber\\
\left[W^{IJ},\,W_{KL}\right]&=&\left[W_{IJ},\,W_{KL}\right]=0\,.\label{comms}
\end{eqnarray}
Closure in $\rE_{7(7)}$ implies the following relations between
the coefficients: \begin{eqnarray}
a_1&=&8\,\alpha\,\,\,;\,\,\,a_2=\alpha=\frac{\gamma}{2}\,\,\,;\,\,\,b_1=24\,\frac{\alpha^2}{\beta}=\frac{b_2}{2}\,,\nonumber\\
c_1&=&-96\,\frac{\alpha^3}{\beta\sigma}\,\,\,;\,\,\,\frac{\lambda}{\sigma}=\frac{6\alpha}{\beta}\,\,\,;\,\,\,\delta=\alpha\,\,\,;\,\,\,\sigma=\alpha\,.
\end{eqnarray}
In particular we can choose $\alpha=\beta=\rho=\sigma=1$ and
$a_3=c_1/a_2$ and eqs. (\ref{comms}) will read:
\begin{eqnarray}
\left[Z_M,\,Z_N\right]&=&\tau_{MN}{}^P\,Z_P+g_{MNPQ}\,W^{PQ}+\tilde{g}\,W_{MN}\,,\nonumber\\
\left[Z_M,\,W^{PQ}\right]&=&2\,\tau_{MR}{}^{[P}\,W^{Q]R}+g_{MM_1M_2M_3}\,\epsilon^{M_1M_2M_3PQRS}\,W_{RS}\,,\nonumber\\
\left[Z_M,\,W_{PQ}\right]&=&
\tau_{PQ}{}^L\,W_{ML}\,\nonumber\\
\left[W^{IJ},\,W^{KL}\right]&=&-3\,
\tau_{I_1I_2}{}^{[K}\,W_{I_3I_4}\epsilon^{L]IJI_1\dots I_4}\,,\nonumber\\
\left[W^{IJ},\,W_{KL}\right]&=&\left[W_{IJ},\,W_{KL}\right]=0\,.\label{comms2}
\end{eqnarray}
 The non--vanishing commutator
between two $W^{IJ}$ generators follows from the embedding of the gauge algebra inside
$\rE_{7(7)}$. In particular it is a consequence of the last commutator in (\ref{come7}), where
$t_M$ are the isometry generators associated with the 7 axions dual to the antisymmetric
tensor fields  $B_I$. If on the other hand these tensors were left undualized, the scalar
manifold would not have had the isometries $t_M$. As a consequence of this, the last
commutator in (\ref{come7}) would vanish and the generators $W^{IJ}$ would be abelian. This is
an example of the phenomenon called \emph{dualization of dualities} discussed in \cite{dd}.
\par Note that the generators $W_{IJ}$ define an abelian ideal
${\frak I}$ inside the gauge algebra ${\frak g}$. If we consider
the quotient algebra
\begin{eqnarray}
\tilde{{\frak g}}&=&{\frak g}/{\frak I}\,,\label{tildeg}
\end{eqnarray}
it has the structure described in (\ref{modg}) and, as we shall see, is also realized in a
subsector of the FDA associated with the theory with undualized antisymmetric tensor fields
\cite{ddf}. Such subsector consists of the massless forms which survive in the effective
theory after the Higgs mechanism between 1-- and 2--forms has taken place \cite{css}.
\subsection{Symplectic frame and the S-S gauging} As previously
pointed out, the second order constraints (\ref{2ord}) guarantee that no more than 28
independent combinations out of $A^M_\mu,\,A_{MN\mu},\,\tilde{A}^{MN}_\mu$ are involved in the
gauging and thus define the actual elementary vector fields of the model. These combinations
are defined by the twist--tensor $\tau$. Let us denote as usual by $r$ the rank of
$\tau_{MN}{}^P$ as a $21\times 7$ matrix. The counting of the elementary vector fields
proceeds as follows. Let us denote now by $A_{P\mu}$ $r$ independent components of the
$A_{MN\mu}$ vectors defined as follows:
\begin{eqnarray}
A_{MN\mu}&=&-2\,\tau_{MN}{}^P\,A_{P\mu}+{\rm
\AA}_{MN\mu}\,,\label{Ad}
\end{eqnarray}
${\rm \AA}_{MN\mu}$ being the remaining $21-r$ components. One can
check that the $r$ vectors $A_{P\mu}$ can always be reabsorbed in
a redefinition of the vectors $\tilde{A}^{MN}_\mu$. Indeed, by
using eqs. (\ref{2cons}) and (\ref{Ad}), the gauge connection
(\ref{conng}) can be rewritten in the following form:
\begin{eqnarray}
\Omega_{\frak{g}\,\mu}&=&A^M_\mu\,Z_M+{\rm \AA}_{MN\mu}\,W^{MN}+\tilde{A}^{\prime MN}_\mu\,
W_{MN}\,,\end{eqnarray} where
\begin{eqnarray}
\tilde{A}^{\prime
MN}_\mu&=&\tilde{A}^{MN}_\mu+A_{P\mu}\epsilon^{MNM_1\dots M_4
P}\,g_{M_1\dots M_4}\,.
\end{eqnarray}
 From the expression of the $W_{MN}$
generators in terms of $t_P$ given in eq. (\ref{Wdd}), we see that only $r$ independent
combinations $\tilde{A}^P_\mu=\tau_{MN}{}^P\,\tilde{A}^{\prime MN}_\mu$ of the 21
$\tilde{A}^{\prime MN}_\mu$, take part in the minimal couplings. Thus the vector fields
actually involved in the gauging sum up to 28 and consist in the $7$ Kaluza--Klein vectors
$A^M_\mu$, the $21-r$ vectors ${\rm \AA}_{MN\mu}$ and the $r$ vectors $\tilde{A}^P_\mu$. These
latter have  Stueckelberg--like couplings to $r$ of the scalars $\tilde{B}^M$, as a
consequence of which they acquire mass through the Higgs mechanism.
\par We may dualize the scalars $\tilde{B}^M$ back to the tensors
$B_{M\mu\nu}$. In this case consistency of the theory requires a
corresponding dualization of the vector fields, associated with a
electric-magnetic duality rotation. The $r$ vectors
$\tilde{A}^P_\mu$ are dualized into the $A_{P\mu}$ components of
the $A_{MN\mu}$ vectors, defined in eq. (\ref{Ad}). These latter
enter the Lagrangian in the following combination with the
antisymmetric tensor fields:
\begin{eqnarray}
dA_{MN}+\tau_{MN}{}^P\,B_P=\tau_{MN}{}^P\,(dA_P+B_P)+d{\rm
\AA}_{MN} \,,
\end{eqnarray}
In the dual theory therefore the $r$ vectors $A_P$ give mass to
$r$ of the tensors $B_P$ by means of an anti--Higgs mechanism
.\par As an example let us consider now the gauging induced by the
following choice on non vanishing components of $\tau,\,g$:
\begin{eqnarray}
\tau_{1m}{}^n\,\,;\,\,\,g_{1mnp}\,\,\,(m,n,p=2,\dots
7)\,.\label{sscomp}
\end{eqnarray}
These components are defined by the branching of the
representations ${\bf 140}_{+3}$ and $\overline{{\bf 35}}_{+5}$
with respect to the ${\rm SL}(6,\mathbb{R})\times {\rm O}(1,1)_1$
subgroup of ${\rm SL}(7,\mathbb{R})$. However the same components
are also defined by the branching with respect to ${\rm
GL}(6,\mathbb{R})\subset E_{6(6)}$ of the $\rE_{6(6)}\times {\rm
O}(1,1)_3$--representation ${\bf 78}_{+3}$ ($\rE_{6(6)}\times {\rm
O}(1,1)_3$ being a subgroup of $\rE_{7(7)}$), contained inside
${\bf 912}$ and defining the embedding tensor for the generalized
S-S gauging \cite{dwst} (the graviphoton originating from the
$D=5\rightarrow D=4$ reduction is given a ${\rm
O}(1,1)_3$--grading $-3$). To show this let us start branching the
${\bf 140}_{+3}$ and $\overline{{\bf 35}}_{+5}$ with respect to
${\rm SL}(6,\mathbb{R})\times {\rm O}(1,1)_1\times {\rm
O}(1,1)_2$:
\begin{eqnarray}
{\bf 140}_{+3}&\rightarrow& \overline{{\bf 6}}_{(-1,+3)}+{\bf
15}_{(-8,+3)}+{\bf 35}_{(+6,+3)}+{\bf 84}_{(-1,+3)}\,\nonumber\\
\overline{{\bf 35}}_{+5}&\rightarrow& {\bf
20}_{(+3,+5)}+\overline{{\bf 15}}_{(-4,+5)}
\end{eqnarray}
The components in (\ref{sscomp}) correspond to the representations
${\bf 35}_{(+6,+3)}$ and ${\bf 20}_{(+3,+5)}$ respectively. The
representations ${\bf 35}$ and ${\bf 20}$ also appear in the
decomposition of the ${\bf 78}$ of $\rE_{6(6)}$ with respect to
its subgroup ${\rm SL}(6,\mathbb{R})$. To prove that the ${\bf
35}_{(+6,+3)}$ and ${\bf 20}_{(+3,+5)}$ belong to the
decomposition of the S-S embedding tensor defined by the ${\bf
78}_{+3}$ we show that they have the right grading with respect to
${\rm O}(1,1)_3$, namely $+3$. The relation between the ${\rm
O}(1,1)_3$--grading $k_3$ and the ${\rm O}(1,1)_{1,2}$--gradings
$k_1,\,k_2$ can be found to be:
\begin{eqnarray}
k_3&=&\frac{1}{7}\,(2\,k_1+3\,k_2)\,.
\end{eqnarray}
Applying this formula to the two representations we find that
their ${\rm O}(1,1)_3$--grading is indeed $+3$.\par
 The gauging induced by the components (\ref{sscomp}) coincides
 with a S-S gauging obtained by reducing the maximal theory in $D=5$
 with a twist matrix given by:
 \begin{eqnarray}
Z_1&=&\tau_{1m}{}^n\,t_n{}^m+g_{1mnp}\,t^{mnp}\in E_{6(6)}\,,
 \end{eqnarray}
 where $t_n{}^m$ are the generators of the ${\rm
 SL}(6,\mathbb{R})$ subgroup of $\rE_{6(6)}$ and $t^{mnp}$ are
 generators in the ${\bf 20}_{+1}$ in the decomposition of the adjoint of $\rE_{6(6)}$
 with respect to ${\rm
 GL}(6,\mathbb{R})$:
 \begin{eqnarray}
{\bf 78}&\rightarrow &{\bf 35}_0+{\bf 1}_0+{\bf 20}_{+1}+{\bf
20}_{-1}+{\bf 1}_{+2}+{\bf 1}_{-2}\,.
 \end{eqnarray}
 Let us comment now on the symplectic frame corresponding to this
 gauging, namely the frame in which all the gauge generators are
 electric. If we arrange the generators $X_u$ in a symplectic
 vector we have:
\begin{eqnarray}
X_u&=&\left(\matrix{Z^1=0\cr Z^m=0\cr  W_{mn}=0 \cr W_{1m}\neq 0\cr Z_1\neq 0\cr Z_m\neq 0\cr
W^{mn}\neq 0\cr W^{1m}=0}\right)\,.
\end{eqnarray}
The symplectic rotation needed to define the right symplectic
frame is effected by switching the ${\bf 6}$ of $W_{1m}$ with the
$\overline{{\bf 6}}$ of the $W^{1m}$. Therefore the new electric
charges transform in the ${\bf 1}+\overline{{\bf 15}}+2\times {\bf
6}$ which complete the ${\bf 27}+{\bf 1}$ of $\rE_{6(6)}$. This
shows that the gauged Lagrangian induced by the torsion/flux
components (\ref{sscomp}) coincides with the generalized S-S
Lagrangian also as far as the symplectic frame is concerned.
\section{The FDA approach}
In this section we give a short resum\'e of the results obtained in \cite{ddf}, to be compared
with the results illustrated in the previous section. The FDA obtained from M-theory
compactification on twisted tori with form-fluxes is given by:
\begin{eqnarray}
dA^I+\frac{1}{2}\,\tau_{JK}{}^I\, A^J\wedge A^K&=&0\nonumber\\
dA_{IJ}+2\,\delta_{[I}^{[L}\,\tau_{J]K}^{M]}\,A^K\wedge
A_{LM}+\frac{1}{2}\,g_{IJKL}\,A^K\wedge A^L+\tau^L_{IJ}\,B_L&=&0\,\nonumber\\
dB_I+\tau_{IL}{}^J \,A^L\wedge
B_J+\frac{1}{6}\,g_{IJKL}\,A^J\wedge A^K\wedge A^L&=&0\,.
\end{eqnarray}
where integrability requires:
\begin{equation}\label{cons}
\tau_{[MN}{}^P\,\tau_{Q]P}{}^R=0\,\,\,\,\,\tau_{[MN}{}^P\,g_{M_1M_2M_3]P}=0\,
\end{equation}
Let us denote by  ${\mathcal A^{(k)}}$ a FDA generated by $p$--forms of degree $p\,\leq k$.
Then a general theorem on the FDAs \cite{sullivan} guarantees that this differential algebra
has a unique decomposition as the  semi--direct sum of two algebras :
\begin{equation}\label{dec}
{\mathcal A^{(k)}} ={\mathcal M^{(k)}}\oplus_s {\mathcal C^{(k)}}
\end{equation}
where the ``contractible algebra'' ${\mathcal C^{(k)}}$ has the
structure
\begin{equation}\label{contr}
d{\mathcal C^{(k)}}\subset {\mathcal C^{(k+1)}}
\end{equation}
while the ``minimal algebra'' ${\mathcal M^{(k)}}$ has the
structure:
\begin{equation}\label{min}
d{\mathcal M^{(k)}}\subset{\mathcal M^{(k)}}\wedge{\mathcal M^{(k)}}
\end{equation}
 An example of
this decomposition has been given in \cite{ddf} in the case of the S-S gauging discussed
earlier and defined by $\tau\equiv\tau_{1i}{}^j=T_i{}^j$. Here we further require
$T_i{}^j=-T^j{}_i$. Defining
\begin{eqnarray}
\hat{B}_i &=&B_i-T^j{}_k\,T^{-1}_i{}^\ell\,A_{\ell j}\wedge A^k+A_{1i}\wedge A^1+
\frac{1}{2}\,T^{-1}_i{}^j\,g_{jk\ell1}\, A^k\wedge A^\ell\,,\nonumber\\
\hat{B}_1 &=&B_1+A_{1i}\wedge A^i\,,\nonumber
\end{eqnarray}
the FDA becomes:
\begin{eqnarray}
dA^1&=&0\,\nonumber\\
dA^i+T^i{}_j\,A^1\wedge A^j&=&0\,\nonumber\\
dA_{ij}+2\,T^k{}_{[i}\, A^1\wedge A_{j]k}+g_{1ijk}\,A^1\wedge A^k&=&0\,\nonumber\\
d\hat{B}_1+T^i{}_j\,A^k\wedge A^j\wedge A_{ki}-\frac{1}{3}\,g_{1ijk} A^i\wedge A^j\wedge A^k&=&0\,\nonumber\\
dA_{1i}+T^j{}_{i}\,\hat{B}_j&=&0\,\,\,\Rightarrow\,\,\,\,d\hat{B}_j=0\,,
\end{eqnarray}
where the first four equations define the minimal algebra ${\mathcal M}^{(2)}$ of which the
first three correspond to the Lie algebra, while the fifth equation defines the contractible
algebra.\par The physical interpretation of the contractible algebra ${\mathcal C}^{(2)}$ is
that it consists in those 2-forms and 1-forms which are involved in the Higgs mechanism: the
tensors $\hat{B}_j$ become massive by eating the vectors $A_{1i}$. As far as the Lie algebra
contained in ${\mathcal M}^{(2)}$ is concerned, it reproduces the structure (\ref{modg}) in
this special example. This latter property is however general and allows us to make the
following statement about the connections between the FDA approach and the gauged supergravity
analysis of the previous section: The gauge Lie algebra contained in ${\mathcal M}^{(2)}$ has
the same structure (\ref{modg}) as the quotient algebra (\ref{tildeg}).

\section{Conclusions}
In the $\rE_{7(7)}$ four--dimensional formulation of M--theory, in
the $\rSL(7,\mathbb{R})$--basis, we have seen that the gauge
algebra,  when the S-S twist $\tau_{IJ}{}^K$ is non--vanishing,
contains both $W^{KL}$ and $W_{KL}$ generators. The total number
of these generators is
 21 while the number of $W_{IJ}$ is bound to be
less or equal to 7, depending on the rank of the $21\times 7$
matrix $\tau_{IJ}{}^K$. When $\tau_{IJ}{}^K= 0$ and $W_{KL}=0$,
the dual algebra (\ref{comms2}) and the original algebra
(\ref{modg}) coincide. This is expected because in this case the
antisymmetric tensors have no magnetic mass terms which couple
them to the gauge generators. However when $\tau_{IJ}{}^K\neq 0$,
depending on its rank $r$ (as a $21\times 7$ matrix), $21-r$
generators are of type $W^{KL}$ and $r$ are of type $W_{KL}$, with
a non--vanishing commutation given in (\ref{comms2}) and depending
on the $\tau$ matrix. This is the dual algebra of the original
M--theory compactification on a twisted torus with fluxes
\cite{df,alt,ddf}. We also note that the gauge vectors
$\tilde{A}_\mu^{IJ}$, corresponding to the $W_{IJ}$ generators,
are $r$ in number and correspond to the vectors which are eaten by
the antisymmetric tensors in the FDA formulation. This result is
not surprising in view of the dynamics of supergravity coupled to
antisymmetric tensor fields studied in \cite{lm}-\cite{dws}.\par
Indeed when massive antisymmetric tensors are dualized into
massive vectors, each of which can be written in terms of a
massless vector plus a scalar field through the Stueckelberg
mechanism, a symplectic rotation occurs for the vector fields
which is equivalent to the appearance of the generators $W_{IJ}$
in the dual gauge algebra structure. Note that this is also
implied by the compatibility with the $\rE_{7(7)}$ symmetry which
is manifest in the dual formulation.\par The two gauge algebra
structures, namely (\ref{modg}) and the dual (\ref{comms2}), just
coincide if one restricts oneself to the smaller gauge algebra
resulting from the quotient by the \emph{abelian ideal} generated
by the additive $r$ generators $W_{IJ}$.\par It is an interesting
problem, in the presence of gauge couplings, to carry out the
dualization of the M--theory Lagrangian in order to show the
equivalence of the two formulations.

\section{Acknowledgements}
S.F. and M.T. would like to thank  G. Dall'Agata and H. Samtleben
for valuable discussions.\par
 Work supported in part by the European Community's Human Potential Program under contract
MRTN-CT-2004-005104 `Constituents, fundamental forces and symmetries of the universe', in
which R. D'A. and M.T.  are associated to Torino University. The work of S.F. has been
supported in part by European Community's Human Potential Program under contract
MRTN-CT-2004-005104 `Constituents, fundamental forces and symmetries of the universe', in
association with INFN Frascati National Laboratories and by D.O.E. grant DE-FG03-91ER40662,
Task C.

 \end{document}